\documentclass[aps, prd, preprint, superscriptaddress, showpacs]{revtex4}
\usepackage{graphicx}

\usepackage{ulem}
\usepackage{epsfig}
\usepackage{color}
\definecolor{My_red}        {cmyk}{0.00, 1.00, 1.00, 0.20}

\newcommand{\bmat}{\left(\begin{array}}
\newcommand{\emat}{\end{array}\right)}
\newcommand{\beq}{\begin{equation}}
\newcommand{\eeq}{\end{equation}}
\newcommand{\wt}{\widetilde}

\def\ra{\rightarrow}

\def\Ld{\Lambda}
\def\ld{\lambda}
\def\f{\frac}

\def\bwt{\begin{widetext}}
\def\ewt{\end{widetext}}
\def\be{\begin{equation}}
\def\ee{\end{equation}}
\def\bea{\begin{eqnarray}}
\def\eea{\end{eqnarray}}
\def\bean{\begin{eqnarray*}}
\def\eean{\end{eqnarray*}}
\def\bary{\begin{array}}
\def\eary{\end{array}}
\def\bit{\begin{itemize}}
\def\eit{\end{itemize}}

\def\ra{\rightarrow}

\def\Ld{\Lambda}

\def\ld{\lambda}

\def\su5u1{SU(5) \times U(1)}
\def\fsu5u1{SU(5) \times U(1)'}
\def\so10{SO(10)}
\def\sq20{SO(10) \times SO(10)}

\def\ra{\rightarrow}

\def\Ld{\Lambda}
\def\ld{\lambda}
\def\f{\frac}

\def\L{\left(}
\def\R{\right)}
\def\bwt{\begin{widetext}}
\def\ewt{\end{widetext}}
\def\be{\begin{equation}}
\def\ee{\end{equation}}
\def\bea{\begin{eqnarray}}
\def\eea{\end{eqnarray}}
\def\bean{\begin{eqnarray*}}
\def\eean{\end{eqnarray*}}
\def\bary{\begin{array}}
\def\eary{\end{array}}
\def\bit{\begin{itemize}}
\def\eit{\end{itemize}}

\def\ra{\rightarrow}

\def\Ld{\Lambda}

\def\ld{\lambda}

\def\su5u1{SU(5) \times U(1)}
\def\fsu5u1{SU(5) \times U(1)'}
\def\so10{SO(10)}
\def\sq20{SO(10) \times SO(10)}

\usepackage[centertags]{amsmath}
\usepackage{amssymb}

\begin{document}

\title{The Minimal Solution to  the $\mu/B_\mu$ Problem in Gauge Mediation}

\author{Zhaofeng Kang}
\email{zhfkang@itp.ac.cn}

\affiliation{Key Laboratory of Frontiers in Theoretical Physics,
             Institute of Theoretical Physics,  Chinese Academy of Sciences,
             Beijing 100190,  P. R. China }

\author{Tianjun Li}
\email{tli@itp.ac.cn}

\affiliation{Key Laboratory of Frontiers in Theoretical Physics,
             Institute of Theoretical Physics,  Chinese Academy of Sciences,
             Beijing 100190,  P. R. China }

\affiliation{George P. and Cynthia W. Mitchell Institute for
             Fundamental Physics,  Texas A$\&$M University,
             College Station,  TX 77843,  USA }

\author{Tao Liu}
\email{tliuphy@itp.ac.cn}

\affiliation{Key Laboratory of Frontiers in Theoretical Physics,
             Institute of Theoretical Physics,  Chinese Academy of Sciences,
             Beijing 100190,  P. R. China }
\author{Jin Min Yang}
\email{jmyang@itp.ac.cn}
\affiliation{Key Laboratory of Frontiers in Theoretical Physics,
             Institute of Theoretical Physics,  Chinese Academy of Sciences,
             Beijing 100190,  P. R. China }

\date{\today}

\begin{abstract}

We provide a minimal solution to the $\mu/B_\mu$ problem in
the gauge mediated supersymmetry breaking by introducing a
Standard Model singlet filed $S$ with a mass around the
messenger scale which couples to the Higgs and messenger fields.
This singlet is nearly supersymmetric and acquires a relatively small
Vacuum Expectation Value (VEV) from its radiatively generated
tadpole term. Consequently, both $\mu$ and $B_\mu$ parameters receive
the tree-level and one-loop contributions, which are comparable
due to the small $S$ VEV. Because there exists a proper cancellation
in such two kinds of contributions to $B_\mu$, we can have a
viable Higgs sector for electroweak symmetry breaking.

\end{abstract}


\maketitle

\section{Introduction}
Gauge mediated supersymmetry breaking (GMSB)~\cite{Dine:1981gu,Dine:1995ag} provides
 an extremely appealing mechanism to solve the flavor problem elegantly in the Minimal
Supersymetric Standard Model (MSSM).
However, the well-known $\mu$ problem, namely the origin of the unique electroweak (EW)-scale
mass parameter $\mu$ in the supersymmetric Higgs mass superpotential
term  $\mu H_u H_d$, cannot be addressed naturally in
the Minimal Gauge Mediation (MGM) framework.
To solve such a $\mu$ problem, the authors in Ref.~\cite{Dvali:1996cu}
propose to couple the two Higgs doublets with the hidden sector messengers and $\mu$
can be generated at one-loop level as follows
\begin{align}
\mu\sim \f{\ld_u\ld_d}{16\pi^2}\f{F}{M},
\end{align}
where $\ld_{u,d}$ are some Yukawa couplings in the hidden sector,
$M$ denotes the messenger mass scale and $\sqrt{|F|}$ is the
effective supersymmetry (SUSY) breaking scale in the hidden sector.
However, in this way the soft Higgs mixing term $B_\mu  H_uH_d$ is
simultaneously generated, leading to the relation $B_\mu \sim \mu
F/M$. Then, for a phenomenologically preferred $\mu\sim
100-1000{\rm\, GeV}$, the $B_\mu$  parameter will be several orders
larger than $\mu^2$, rendering the electroweak symmetry breaking
(EWSB) problematic. This is the so-called $\mu/B_\mu $ problem in
the GMSB.

Various attempts have been made to solve this $\mu/B_\mu $ problem
~\cite{Hall:2002up,Giudice:2007ca,Roy:2007nz,Murayama:2007ge,Csaki:2008sr,Komargodski:2008ax}.
For example, one may use a complicate dynamical structure to generate $\mu$ at
one loop  whereas $B_\mu $ does not appear until
two loops~\cite{Dvali:1996cu,Hall:2002up,Giudice:2007ca} or
even vanishes at the messenger boundary~\cite{Komargodski:2008ax}.
One can also work in the next-to-minimal supersymmetric standard model (NMSSM)~\cite{NMSSM}
to generate the $\mu$ parameter from the Vacuum Expectation Value (VEV)
 of a dynamical field, which, however,
is proved to be difficult to generate a large enough $\mu$ in the
pure GMSB~\cite{Dine:1993yw,deGouvea:1997cx} (This problem could be
solved through coupling the singlet field to hidden
sectors~\cite{Ellwanger:2008py}). In addition, the renormalization
of the strong dynamics in the hidden sector may affect differently
on the $\mu$ and $B_\mu $
parameters~\cite{Murayama:2007ge,Csaki:2008sr}, but such models with
strong dynamics suffer from the incalculable problem. An interesting
observation was made in Ref.~\cite{Csaki:2008sr}, which found that a
desired Higgs sector with successful EWSB does not always need a
small $B_\mu$ while a large $B_\mu$ together with a correspondingly
large $m_{H_d}^2$ can also make EWSB work. Such a scenario does not
require new fine-tuning and has distinct phenomenology
~\cite{DeSimone:2011va}.

Based on the above progresses, we in this paper attempt to explore a
minimal solution to the $\mu/B_\mu$ problem. For this purpose, we
try to introduce the minimal degrees of freedom with the minimal new
scales. But we require the calculability and perturbativity up to
the Grand Unified Theory (GUT) scale. We find that
we can make it by extending the
minimal gauge mediation with a Standard Model (SM) singlet field
$S$ around the messenger scale which couples to the
Higgs and messenger fields. The heavy singlet $S$ is nearly
supersymmetric, {\it i.e.}, the mass of its scalar component is almost
equal to its fermionic component.
And it obtains a relatively small VEV due to the radiatively
generated tadpole term. Therefore, both
$\mu$ and $ B_\mu$ parameters can receive the tree-level and one-loop
contributions that are comparable due to the small
$S$ VEV. Our crucial observation is that such two kinds of
contributions to  $B_\mu$ can allow for a proper cancellation.
Thus, a viable Higgs sector for EWSB can be realized.

The paper is organized as follows. In Section II we present the minimal solution.
In Section III we investigate the EWSB from the generated Higgs sector and
comment on the supersymmetric CP problem. The Conclusion is made in Section IV.

\section{The Minimal Solution to the $\mu/B_\mu $ Problem}
\subsection{Minimal Gauge Mediation }
We start from the superpotential in
the minimal gauge mediation
\begin{align}
W_{h}=X\phi\bar\phi
\end{align}
with the spurion field $X=M+\theta^2 F$ and $N$ pairs of messengers  $(\phi,\bar\phi)$
which fill the ($5,\bar 5$) representations of $SU(5)$ gauge group.
The supersymmetry breaking
soft mass spectrum at the messenger scale is given by
\begin{align}\label{PGMSB}
M_{a}\simeq& N\frac{\alpha_a}{4\pi}\Ld, \\
m_{\wt f}^2\simeq&2N
\sum_aC_a^f\L\frac{\alpha_a}{4\pi}\R^2 \Ld^2,
\end{align}
where $\Ld\equiv F/M$, $a=1,~2,~3$, $\alpha_1$, $\alpha_2$ and
$\alpha_3$ are respectively the gauge couplings for $U(1)_Y$,
$SU(2)_L$ and $SU(3)_C$, and $C_a^f$ denote the quadratic Casimir
invariant for the particle $\wt f$ with respect to three SM gauge
groups.

\subsection{Higgs-Messenger Mixings and Tree-Level $\mu/B_\mu $ terms}

The Higgs doublets $H_u$ and $H_d$ just assemble a pair of messengers:
they have a supersymmetric mass term $\mu H_uH_d$ as well as
a SUSY-breaking term $B_\mu  H_uH_d$.
Naively, one expects that coupling them directly  to the Goldstino field  $X$ via
$\ld  X H_uH_d$ may solve the $\mu/B_\mu -$problem, provided that $\mu=\ld  M\sim{\cal O}(m_Z)$.
However, this does not work because the robust relation $B_\mu /\mu^2=\Ld/\mu$ still holds
even if $\ld $ can be as small as we want.

In order to see how the Higgs fields can feel the SUSY-breaking effect through small
Higgs-messenger mixings, we consider an example
\begin{align}
W_{HV}=X\phi\bar\phi+\L v_1H_u\phi_L+ v_2H_d\bar\phi_L\R,
\end{align}
where the $SU(2)_L$ doublet components of the messengers are denoted
as $(\phi_L,\bar\phi_L)$, and  $v_1$ and $v_2$ are the introduced new scales.
Integrating  out the messengers $(\phi_L,\bar\phi_L)$ at
tree level, we obtain $\mu$ and $B_\mu $ as follows
\begin{align}\label{TmuB}
\mu=&-  \frac{v_1v_2}{M},\cr
B_\mu =&- \frac{v_1v_2 }{M^2}\times F=\mu \Ld,
\end{align}
which are valid at small mixing limits $v_{1,2}\ll M$ ( $v_{1,2}$ are
naturally small because in the UV completed models they are induced by radiative corrections).
We see that the  $\mu/B_\mu $ problem remains unsolved due to the relation $B_\mu=\Ld \mu$.

Some comments on the above results are in order.
First, the SUSY breaking parameter $B_\mu $ is proportional to the extent
of SUSY-breaking $F/M^2$ while the supersymmetric parameter $\mu$ is not
related to SUSY-breaking.
Second, the above result is easy to understand. Rotating the Higgs doublets
and messenger doublets to the mass eigenstates (with one eigenvalue $\mu$),
we obtain an effective coupling
 \begin{align}
W_{HV}\supset\L F_{\phi u}F_{\bar \phi d}\R X H_uH_d,
\end{align}
where we use the same notation for the fields before and after rotation.
Here  $F_{\phi_{u,d}}(v_{1},v_2,M)$ are the light doublet fractions contained  in the
messengers. We just recover the  model at the beginning of this section by setting
$\ld =F_{\phi u}F_{\bar \phi d}$.

By the way, in the above proposal some ad hoc new scales $v_{1,2}$ have to be introduced
by hand.
So it seems that we just trade one problem with another.
In our new attempt we will try to dynamically generate such a new scale from
the VEV of a new SM singlet field  $S$. Consequently, additional contributions
 arise from one loop, which can naturally (when we consider the complete dynamics of $S$)
reduce the large tree-level $B_\mu $.

\subsection{Dynamical Mechanism and One-Loop Contributions to $\mu/B_\mu $}

In order to overcome the tree-level relation between $\mu$ and $B_\mu$,
we dynamically generate the scales $v_{1,2}$ by choosing the superpotential
as follows
\begin{align}\label{model}
 W=\ld_u SH_u\phi_L+\ld_d
SH_d\bar\phi_L+\ld  X\phi\bar\phi+f(S).
\end{align}
where $\ld_u$, $\ld_d$, and $\ld $ are Yukawa couplings. Also,
 $f(S)$ denotes the complete dynamics of $S$ which generates a VEV for
$S$ at $\langle S\rangle \equiv v$ (the complete form for $f(S)$ will be
specified in the next Section). Then we have $v_{1,2}=\ld_{1,2}v$.
The key dynamical information of $S$ can be parameterized as
\begin{align}\label{modelv}
W\supset\ld_u{\cal S}H_u\phi_L+\ld_d{\cal S}H_d\bar\phi_L +\ld
X\phi\bar\phi+\frac{1}{2}M_{{S}}{\cal S}^2+\ld_S{\cal S}^3,
\end{align}
where we have made the shift $S\rightarrow v+{\cal S}$ and simply assumed
the effective theory for $S$ is a polynomial.
Note that in our approach $S$ is  supersymmetric (we will turn back to this point later),
 which is contrary to the cases discussed in Ref.~\cite{Evans:2010kd,DeSimone:2011va}.
This will lead to much different results between our work and previous studies.
\begin{figure}[htb]
\begin{center}
\includegraphics[width=4.5in]{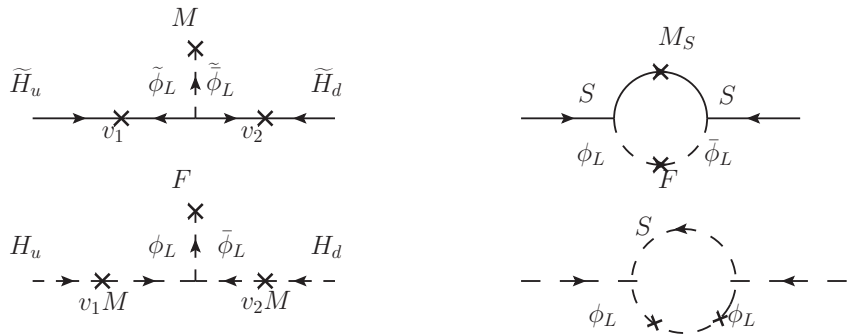}
\end{center}
\vspace*{-1cm}
\caption{\label{cascade} The generation of $\mu/B_\mu$ at tree and loop level.}
\end{figure}
A crucial observation from Eq.~(\ref{modelv}) is that with a
dynamical $S$ propagating in the loop, there will be one-loop
contributions to the $\mu/B_\mu $ parameters, which are proportional
to $F$. Now we calculate the contributions from the one-loop diagrams
shown in Fig.~1. We start from  Eq.~(\ref{modelv}) and neglect the
small Higgs-messenger mixings. Note that the dimensionless parameter
$\ld_S$ is irrelevant because it does not enter into the one-loop
diagrams. The analytical results are given by
\begin{align}
\Delta\mu=& -\frac{\ld_u\ld_d}{16\pi^2}f_1(x)\Ld, \\
\Delta
B_\mu =&
-\frac{\ld_u\ld_d}{16\pi^2}f_2(x)\Ld^2,
\end{align}
where $x\equiv {M_S}/{M}$, and the functions $f_{1,2}(x)$ are defined
by~\footnote{Our calculations confirm the recent results in Ref.~\cite{DeSimone:2011va},
where a method  based on the effective
K$\ddot{a}$hler potential was used.}
\begin{align}\label{f12}
f_1(x)=&\frac{x}{(x^2-1)^2}(x^2\log x^2-x^2+1),\cr
f_2(x)=&\frac{x}{(x^2-1)^3}(-2x^2\log x^2+x^4-1).
\end{align}
Notice that these one-loop contributions are proportional to $M_{{\cal S}}$. This can be
easily understood because turning off the mass term of ${\cal S}$ will render
the model a global $U(1)$ symmetry under which $\cal S$ is charged, aside from the
irrelevant term $\ld_S{\cal S}^3$.
Now the total values of $\mu$ and $B_\mu $ are given by
\begin{align}\label{totalmub}
\mu=&-\ld_u\ld_d\left[ f_v+{f_1(x)}\right] \f{\Ld}{16\pi^2},\\
B_\mu =&\mu \f{  f_v+f_2(x) }{ f_v+f_1(x) }\Ld,
\end{align}
where $f_v= 16\pi^2 v^2/F$. Since $B_\mu $ receives contributions at
both tree level and one loop, these contributions may cancel each
other to some extent for a negative $F$. In particular, if the $S$
VEV is small, $f_v$ and $f_2(x)$ can be comparable.

Through the Yukawa interactions the soft masses of $H_u$ and $H_d$
can also get the one-loop (with $\cal S$ running in the loops)
non-holomorphic contributions, given at the leading order by
\begin{align}
\Delta m_{H_u}^2=&\frac{\ld_u^2}{16\pi^2}\Ld^2g(x)
,\quad
\Delta  m_{H_d}^2=\frac{\ld_d^2}{16\pi^2}\Ld^2g(x),
\end{align}
where $g(x)$ is defined as
\begin{align}
g(x)=\f{x^2}{(x^2-1)^3}\left[(1+x^2)\ln x^2+2(1-x^2)\right].
\end{align}
Note that there are also tree-level soft masses for $H_{u,d}$ which
are proportion to $v^4F^2/M^6$ and much smaller than
the above one-loop contributions, so they can be omitted safely.
It is well known that in the Yukawa-deflected
models~\cite{Delgado:2007rz,Kang:2010ye,Kang:2010mh} such one-loop
soft masses vanish at the leading order of SUSY-breaking and they
are generated only at two-loop level through the wave function
renormalization~\cite{Giudice:1997ni,ArkaniHamed:1998kj}. Here in
our model the one-loop contributions survive by virtue of the fact
that $S$ does not couple to the (single) spurion field $X$ and thus
avoid the accidental cancellation (we will turn to this point
later). In addition, the trilinear soft terms  $A_u(h_u)_i \wt
Q_iH_u\wt U^c_i$ are also generated
\begin{equation}
A_{u,d}= \f{\ld_{u,d}^2}{16\pi^2}h(x)\Ld,
\end{equation}
with the function $h(x)$ defined as
\begin{equation}
h(x) = \f{1}{(x^2-1)^2}\left[(x^2-1)-x^2\log x^2\right].
\end{equation}
As shown later, these trilinear soft
terms do not play a significant role in our discussions.

\begin{figure}[htb]
\begin{center}
\includegraphics[width=3.5in]{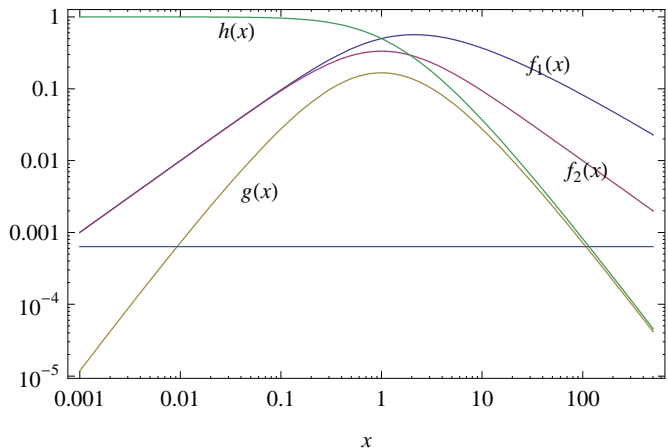}
\end{center}
\vspace*{-1cm} \caption{\label{FIG2} Plot of the functions $f_1(x)$,
$f_2(x)$, $g(x)$ and $|h(x)|$.}
\end{figure}

In the above results we defined four functions.
While the function $h(x)$ is negative, the other three
functions $f_{1,2}(x)$ and $g(x)$,
relevant to the Higgs parameters, are semi-positive.
Especially, $h(x)$ is not proportional to $x$, and thus it is
the most important function in the small $x$ region.
Such properties may be useful, for example, may help to lift up the
lightest CP-even Higgs boson mass.
Note that for $f_{1,2}(x)$ and $g(x)$
there is a relation
\begin{align}\label{GX}
g(x)=\f{f_1(x)-f_2(x)}{x}
\end{align}
and further they satisfy
\begin{align}\label{FUNS}
0<g(x)< f_2(x)< f_1(x)<1.
\end{align}
For $x\ll1$, we have $g(x)\ll f_1(x)\simeq f_2(x)\ll 1$, it is easy
to see that  $B_\mu \sim \Ld \mu $ from Eq.~(\ref{totalmub}), so in
this paper we will turn to the regions $x > 0.2$;  while for a large
$x$ we have the hierarchy $g(x)\ll f_2(x)\ll f_1(x)\ll1 $, which is
helpful to solve the $\mu/B_\mu-$problem. These properties are shown
Fig.~\ref{FIG2}.

In the Higgs sector there are four effective free parameters: $\ld_u$,  $\ld_d$, $f_v$ and $x$.
Especially $f_v$ plays an important role in the cancellation between the tree-level
 and one-loop contributions
of $B_\mu$ because it only appears at tree level.
 It is convenient to define
 \begin{align}\label{aB}
f_v+f_2(x)\equiv a_Bg(x),
\end{align}
and hereafter we use $a_B$ to replace the parameter $f_v$. If
$f_v>0$, we obtain $ a_B >f_2(x)/g(x)$, which slowly varies with
$x$, reaching a minimal value around 2 when $x$ is about 1 (see
Fig.~\ref{FIG2}). By contrast, if $ a_B<2$, we should have the
cancellation between $f_v$ and $f_2(x)$. Naively, a very small $a_B$
implies a severe tuning and w will turn to this problem later. Now
the parameters $\mu$, $B_\mu$, $\Delta m_{H_u}^2$ and $\Delta
m_{H_d}^2$ can be written as
\begin{align}\label{LBMU}
&\mu=-g(x)\ld_u\ld_d (x+  a_B)\f{\Ld}{16\pi^2},\quad
B_\mu=\f{a_B}{x+a_B}\mu\Ld,\cr &\Delta m_{H_u}^2\simeq -\f{1}{|r|(x+
a_B)}\mu\Ld,\quad \Delta m_{H_d}^2\simeq  -\f{|r|}{x+  a_B}\mu\Ld,
\end{align}
where $r\equiv \ld_d/\ld_u$,  and in our convention $B_\mu>0$. Here
we see that $g(x)$ can be absorbed into the Yukawa couplings via the
rescaling $\ld_{u,d}\ra g^{1/2}(x)\ld_{u,d}$. Although the value of
$g(x)$ is irrelevant to our solution,  it affects the perturbativity
of the couplings. As shown in Fig.~\ref{FIG2}, $g^{1/2}(x)\lesssim
0.3$. From the above expressions we see that for a large $x$, only
the parameter $\mu$ is enhanced while the other three parameters
including  $B_\mu$ are suppressed. So it is possible to suppress
$B_\mu$ relative to  $\mu$, and then the $\mu/B_\mu$ problem is
solved.

Note that with only the one-loop contributions the parameters
$B_\mu$, $m_{H_u}^2$ and $m_{H_d}^2$ are usually correlated
undesirably, leading to an uncontrollably large $B_\mu$. As
discussed in \cite{DeSimone:2011va}, in order to break such a
correlation, some new fields and scales have to be introduced.
However, in our proposed minimal solution, since  $B_\mu$ receives
both the tree-level and one-loop contributions and these two kinds of
contributions can cancel each other to some extent
if they are comparable, such a correlation can be
relaxed, albeit with reasonable fine-tuning.

\subsection{The Complete  Model}\label{MDMODL}

In the above we have presented the viable minimal solution to the
$\mu/B_\mu$ problem through a proper cancellation between the tree-level and
one-loop contributions of $B_{\mu}$.
In this solution, we have a key assumption that
the hidden singlet $S$ develops a small VEV $v\ll M$. Now we
discuss the generation  of this small scale.

Since $S$ is a singlet, at the
renormalizable level we can in general expect the supersymmetric  effective theory
of $S$ takes a form of
\begin{align}
W=&aM^2S-\frac{M_SS^2}{2}+\frac{\ld_SS^3}{3},
\end{align}
where $a$ and $\ld_S$ are dimensionless parameters, and $M_S$ is a mass
parameter. Of course,
some term(s) can be forbidden by imposing some proper symmetry. In
the following we  discuss a simple case with $\ld_S=0$. Then the
$F-$flatness of $S$ determines a VEV: $v=aM^2/M_S$. On the other
hand, for the cancellation between tree and loop contributions we
have
\begin{align}\label{TADPLOE}
f_v\sim -f_2(x)\Rightarrow a\sim \f{x\sqrt{f_2(x)}}{4\pi}\L\f{\Ld}{M}\R^{1/2}.
\end{align}
In our interested region $f_2(x)\sim {\cal O}(10^{-2})- {\cal O}(10^{-1})$, the value of
$a$ should be highly suppressed especially when we have a large  messenger scale.

Motivated by an very small $a$, here we give a simple  realization.
 It is natural  to conjecture that a small VEV of $S$ is driven by a small tadpole term,
generated by radiative mechanism.
So we consider a simple model with the following superpotential in
the hidden sector
\begin{align}\label{SMODEL}
 W_{hidden}=&\ld_u SH_u\phi_L+\ld_d
SH_d\bar\phi_L+\kappa S\phi\bar\phi
+\frac{1}{2}M_SS^2+X\phi\bar\phi.
\end{align}
It conserves a discrete $Z_2$ symmetry
\begin{align}\label{}
S\ra -S,\quad X\ra -X,\cr
\phi\ra -\phi,\quad H_d\ra -H_d,
\end{align}
with other fields invariant under this symmetry.
With such a $Z_2$ symmetry the bare term $\mu H_uH_d$ is forbidden.
In this model the singlet $S$ (as the spurion field) couples to the
messenger pair directly and consequently a term $\int
d^4\theta \epsilon (SX^\dagger+XS^\dagger) $ is generated.
Replacing the spurion field $F-$term VEV, we  get the  small tadpole term
$(\epsilon F)S+h.c.$, with $\epsilon$ given by
\begin{align}\label{TADPLOE1}
\epsilon \simeq N_f\frac{\kappa}{16\pi^2}\log\frac{\Lambda_{\rm UV}}{M} ,
\end{align}
where $\Lambda_{\rm UV}$ is the scale at the limit of zero mixing,
$N_f$ is the number of the fields running in the loop ($N_f=5$ in
the minimal case). Naturally it is expected that $\epsilon\sim{\cal
O}(0.1)$. Now the effective parameter $a=\epsilon
{F}/{M^2}=\epsilon\Ld/M$. From the constraint in
Eq.~(\ref{TADPLOE}), we require $\epsilon\sim x \L
f_2(x)\f{M}{\Ld}\R^{1/2}/{4\pi}$. Apparently, in this mechanism $M$
should be relatively light, say $\lesssim 10^9$ GeV, so as to be
consistent with Eq.~(\ref{TADPLOE1}). Otherwise, the generated
tadpole may not enable $f_v$ to cancel $f_2(x)$.

Noet that although only the tadpole mechanism for the generation of  the $S$ VEV
is discussed above (we believe it is the simplest way without involving
other unknown dynamics), one can consider other possibilities,
$e.g.$, $a=0,~\ld_S\neq0$, and $M_S$ originates from the $U(1)_R$ symmetry breaking.
There exists a vacuum with $F_S=0$ where $S$ obtains a small VEV.

\section{Discussions on Low Energy Phenomenology}

In this Section we discuss some low energy phenomenology of our model.
We will first discuss whether or not our model can give successful EWSB.
Then we will make a brief comment on the supersymmetric CP-problem,
which cannot be solved automatically in our model.

\subsection{Electroweak Symmetry Breaking}
\subsubsection{Requirement of electroweak symmetry breaking}
In the MSSM a viable Higgs sector triggering successful EWSB
implies that at the EW-scale they satisfy the following two
minimizing equations
\begin{align}\label{EWSOL1}
\sin2\beta=&\frac{2B_\mu }{m_{H_u}^2+m_{H_d}^{2}+2\mu^2},\\
\frac{m_Z^2}{2}=&-\mu^2+\frac{m_{H_d}^2-\tan^2\beta\,
m_{H_u}^2}{\tan^2\beta-1},\label{EWSOL2}
\end{align}
where $\tan\beta$ is the ratio of Higgs VEVs, and $m_Z$ is the
$Z$ boson mass. All the parameters take values at the EW scale.
Among them, $m_{H_u}^2(m_Z)$ is driven negative  by the  stop RGE effect, and leads to
the well-known radiative EWSB
\begin{align}\label{mHU2}
m_{H_u}^2(m_Z)=\Delta m_{H_u}^2+m_{H_u}^2(M)-\f{3\alpha_t}{\pi}m_{\wt t}^2\log\f{M}{m_{\wt t}}~,~\,
\end{align}
where $\alpha_t$ is the top quark Yukawa coupling. Also,
 the second term in the r.h.s is the pure GMSB contribution given in
Eq.~(\ref{PGMSB}), and the stop mass $m_{\wt t}$ is defined as the geometric mean
of the two stop masses.
In addition, the EW-scale $\mu$ and $B_\mu $ can be approximated to be the value
at the boundary since they are not quite sensitive to the RGE evolutions.
At the leading order their RGEs are given by~\cite{Martin:1997ns}
\begin{align}\label{mubmuRGE}
\f{d\mu}{dt}\simeq&\f{3\L\alpha_t-\alpha_2\R}{4\pi}\mu,
\\
\f{dB_\mu }{dt}\simeq&\f{3\L\alpha_t-\alpha_2\R}{4\pi}B_\mu +\L\f{6\alpha_t}{4\pi}A_u+\f{6\alpha_b}{4\pi}A_d+\f{6\alpha_2}{4\pi}M_2\R\mu,\label{BMURGE}
\end{align}
where $t\equiv\log\f{ Q}{Q_0}$ with $Q$ the running scale and $Q_0$
the referred scale, $A_{u,d}$ are the trilinear soft terms of the
stops and sbottoms, and $M_2$ is the $SU(2)_L$ gaugino mass. As shown later,
in the viable parameter space $\mu$ is small while $B_\mu$ is large,
the RGE correction can be ignored in our analysis.

Note that in the MSSM we need a heavy stop $m_{\wt t}\sim{\cal O}(1)$ TeV to lift up
 the lightest CP-even Higgs boson $h$ mass to satisfy the LEP bound $m_h>114.4$ GeV.
So in the pure GMSB $-m_{H_u}^2(m_Z)$ will be of the stop mass scale (see Eq.\ref{mHU2})
which is far above $m_Z$ and thus implies a fine-tuning between $\mu^2$ and $m_{H_u}^2$
(the tuning extent is roughly measured by $\eta\equiv m_Z^2/m_{\wt t}^2$) to satisfy
Eq.~(\ref{EWSOL2}). This is the so-called little hierarchy problem of the MSSM.

To satisfy the two equations in Eqs.(\ref{EWSOL1}) and (\ref{EWSOL2}),
the Higgs parameters can be classified into the following
three scenarios with different phenomenological consequences
(a similar but more complicate classification can be found in Ref.~\cite{Zheng:2011cf}):
 \begin{description}
  \item[I:] Large $\mu$  and large $|m_{H_u}^2(m_Z)|$.
In the ordinary GMSB like MGM, $\mu$  is an input parameter which is tuned to
cancel the large value of $m_{H_u}^2(m_Z)$
\begin{align}\label{largemu}
-m_{H_u}^2(m_Z)-\mu^2\simeq m_Z^2/2.
\end{align}
In this scenario the lightest neutralino is bino-like,
while the Higgsino-like neutralino/chargino  are as heavy as $\sim {\cal O}(1)$ TeV.
As for the Higgs spectrum, apart from a light Higgs $h$,
other Higgs states are nearly degenerate and quite heavy
\begin{align}
m_A\simeq m_{H^0}\simeq m_{H^\pm}\simeq \max\{|\mu|,m_{H_d}(m_Z)\}\gtrsim{\cal O}(1 \rm \,TeV).
\end{align}
So in this scenario we may only have a light Higgs boson at the LHC.

  \item[II:] Small $\mu$ and small $m_{H_d}^2(m_Z)$.
  In a large class of models with dynamical $\mu/B_\mu$ origin,
  due to the Higgs-messenger direct coupling,
  there are extra (probably large by virtue of being
  generated at one-loop level) contributions to Higgs soft masses at the boundary.
Consequently, it is likely that such extra positive contribution
$\Delta m_{H_u}^2$ may cancel a large portion  of the stop RGE
contribution, leading to $m_{H_u}^2(m_Z)\sim -{\cal O}(m_Z^2)$.
Note that this does not mean that the fine-tuning problem is
solved because we just shift the tuning to a new place, {\it i.e.}, a
unnatural cancellation between $\Delta m_{H_u}^2$ and the stop RGE
contribution. Evidently,  $\mu$ should be around $m_Z$ as well. If
further $m_{H_d}^2(m_Z)$ is also small ($e.g.$, determined by pure
GMSB or even smaller for some other reasons), then $B_\mu $ must
be small in light of Eq.~(\ref{EWSOL1}). In realistic models such
a scenario may incur new fine-tuning. This scenario is
characterized by no very heavy states in the EW-sector, indicating
a more interesting discovery potential at the early LHC run.

\item[III:] Small $\mu$ and large $m_{H_d}^2(m_Z)$.
For a large $m_{H_d}^2(m_Z)$ (maybe due to the large new soft term $\Delta m_{H_d}^2$),
Eq.~(\ref{EWSOL1}) can be satisfied even with a large $B_\mu $~\cite{Csaki:2008sr}.
This can be realized in the lopsided GMSB ~\cite{DeSimone:2011va}, which
dose not incur new fine-tuning but in general requires a  hierarchy $\ld_u\ll\ld_d$.
Moreover, this kind of models also predict a heavy Higgs spectrum.
 \end{description}
\subsubsection{Realization of electroweak symmetry breaking in our model}
Now we show that our minimal model can realize electroweak symmetry breaking
via scenario-III listed above without too much difficulty. First we check the
constraints on the Higgs parameters:
\begin{itemize}
  \item  The
 collider gives a lower bound on the chargino mass, which requires $|\mu|>100$ GeV.
And $\Delta m_{H_u}^2$ should roughly be below (1 TeV)$^2$ so as to
keep Eq.~(\ref{mHU2}) negative to trigger EWSB (actually, a negative
   $m_{H_u}^2(m_Z)$ helps to trigger EWSB but is not a necessary condition).
Then we have
\begin{align}\label{LBMUC}
16\pi^2\f{10^2\rm\,GeV}{\Ld}\f{1}{|(x+  a_B)r|}\lesssim&
g(x){\ld_u^2}\lesssim 16\pi^2 \f{3\alpha_t}{\pi}\f{m_{\wt
t}^2}{\Ld^2}\log\f{M}{m_{\wt t}}\equiv  I_u,
\end{align}
As an illustration we take $m_{\wt t}=1.5$ TeV and $\Ld=100$ TeV, which gives
an upper bound $ I_u\simeq0.08$. For the scale M we take $M=M_{\rm GUT}$
for a general analysis in this subsection and a smaller $M$ such as
$10^9$ GeV discussed previously will not change our conclusion.
And roughly we can get $|( x+ a_B)r|\gtrsim {\cal O}(1).$

Note that the above constraint also implies that the  RGE effect in
$\mu/B\mu$ (see Eq.~\ref{mubmuRGE}) is small. Since $|\Delta
m_{H_u}^2|=|A_u \Ld g(x)/h(x)|\lesssim 10^6\rm\,GeV^2$, we have
$A_u\sim {\cal O}(10)$ GeV in the interesting region of $x>1$. Then
the RGE of $B_\mu$ is dominated by gaugino effect and thus is small.
So in our model $B_\mu$ is dominated by its  boundary value.

\item In the MGM, the requirement of $U(1)_Y$ anomaly free ensures the traceless of
the sparticles masses at the scale $M$, namely ${\cal S}_Y={\rm
Tr}(Y_{\wt f}m_{\wt f}^2)=0$, with tracing over all $U(1)_Y$-charged
sparticles. However, this sum rule is violated due to the generation
of new soft masses from Yukawa interactions, $e.g.,$ ${\cal
S}_Y\simeq -\Delta m_{H_d}^2$. Then it induces an effective
Fayet-Iliopoulos (FI) term for the $U(1)_Y$-charged fields via RGE
running ~\cite{Martin:1997ns}:
\begin{align}\label{}
\delta m_{\wt f}^{2\rm (FI)}\simeq-\f{3Y_{\wt f}\alpha_1 {\cal S}_Y}{10\pi}\log\f{M}{\mu_0}.
\end{align}
For the fields carrying negative hypercharge (for example the
left-handed sleptons), their soft masses will be decreased
considerably (recall that $\Delta m_{H_d}^2$ is generated at
one loop). So to avoid tachyonic sleptons, $\ld_d$ is upper bounded by
\begin{align}\label{LBMU2}
 g(x)\ld_d^2\lesssim \f{5\alpha_2^2}{8\pi\alpha_1}\f{16\pi^2}{L_M}\equiv  I_d,
\end{align}
 where $L_M= \log\f{M}{m_{\wt t}}$. Roughly, $I_d\simeq 0.1$ and thus  $\Delta m_{H_d}$
is below several TeVs.

\item Since both $\ld_{u}$ and $\ld_{d}$ are bounded above, then for a fixed $x$
we get an upper bound on  $\mu$
    \begin{align}\label{sin2beta}
|\mu|\lesssim \min\{ I_u|r|,\f{ I_d}{|r|}\}|x+ a_B
|\f{\Ld}{16\pi^2}{\rm\,}.
\end{align}
>From  this naive estimation,  $|\mu|$ gets its maximal value for
$r\simeq \sqrt{ I_d/ I_u}$
     \begin{align}\label{Maxmu}
|\mu|\lesssim  (x+  a_B)\sqrt{ I_d  I_u} \f{\Ld}{16\pi^2}{\rm\,}.
\end{align}
The bound is saturated if and only if $g(x)\ld_{u,d}^2=I_{u,d}$.
Taking $\ld_{u,d}<4\pi$ (purterbativity) into consideration, then
$g(x)>\max\{I_u,I_d\}/(4\pi)^2$ and thus $x\sim 10^{-2}-10^2$. And
it allows for a maximal $|\mu|\simeq 10$ TeV when $x\simeq10^2$.
Note that if the messenger scale is orders lower, then
$\ld_{u,d}\lesssim1$ for keeping perturbative until GUT scale and also
$I_u$ will be reduced, so the maximal value of $\mu$ will be
smaller.
\end{itemize}

With the constraints in Eqs.~(\ref{LBMUC}), (\ref{LBMU2}) and (\ref{Maxmu}),
we now investigate the EWSB in our model.
First let us consider a TeV scale $|\mu|$ in Scenario-I.
If $x\sim{\cal O}(1)$ ($x+a_B\sim {\cal O }(1)$),
as shown from Eq.~(\ref{LBMU}), we can have a TeV scale $|\mu|$.
But then, independent of the value of $|r|$, we run into either an excessively
large $\Delta m_{H_d}^2$ or $\Delta m_{H_u}^2$. Therefore the case with $x\sim{\cal O}(1)$
is in contradiction with Eqs.~(\ref{LBMUC}) and (\ref{LBMU2}).
If $x$ is large, {\it e.g.}, $x\sim 50$, then from Fig.~2 we see a suppressed
$g(x)\sim1 0^{-3}$. In this case, in order to have a TeV-scale $|\mu|$, we require
large $|\ld_u\ld_d|$ from Eq.~(\ref{largemu})
\begin{align}\label{}
|\ld_u\ld_d|\simeq\f{1}{g(x)(x+ a_B)}\L16\pi^2\f{m_{\wt t}}{\Ld}\R\L\f{3\alpha_t}{\pi}L_M\R^{1/2}
=\f{4\pi \sqrt{I_u}}{g(x)x},
\end{align}
where $\sqrt{I_u}\simeq0.3$ and thus $|\ld_u\ld_d|\sim {\cal O}(10)$.
This can be only allowed by GUT scale messengers. However, in our model
the messenger scale is below $10^9$ GeV, as discussed below Eq.~(\ref{TADPLOE1}).

In addition, from Eqs.~(\ref{EWSOL1}) and (\ref{EWSOL2}) we obtain
 \begin{align}\label{}
1/\tan\beta\simeq \f{a_B}{|r|+(x+a_B) \sqrt{I_u}/4\pi}.
\end{align}
It is well known that in the MSSM a large $\tan\beta$ is favored to
push up the lightest Higgs boson mass. So we require
$\tan\beta\gtrsim 10$ and $a_B\sim {\cal O}(0.1)$.

Next consider the small $\mu$ scenarios: Scenario-II and
Scenario-III. Again we start from the Higgs soft masses
in Eq.~(\ref{LBMU}). First of all, recall that a TeV-scale
 $\Delta m_{H_u}^2$ is required to cancel the stop RGE contribution,
namely $1/(|r|(x+a_B))\sim{\cal O}(0.1)$ . If we further want to
make $\Delta m_{H_d}^2$ far below TeV scale, then typically we need
$x\sim 100$ and $|r|\sim 0.1$ . For a moderate $B_\mu$, an small
$a_B$ of order ${\cal O}(0.01)$ is needed. At the same time due to
the fact that $\ld_u\ld_d \sim 1$ to obtain a proper $\mu$, a large
$\ld_u$ is unavoidable. So it is beyond our minimal dynamical model
neither.

Next we turn to the most attractive case in which $\mu$ is small but
$\Delta m_{H_d}$ is at TeV scale, namely the lopsided scenario.
In this case the EW-breaking equations in Eqs.~(\ref{EWSOL1}) and (\ref{EWSOL2})
can be approximated as follows
 \begin{align}\label{sin2beta}
1/\tan\beta\simeq&\f{B_\mu }{m_{H_d}^2(m_Z)}\left[ 1+{\cal O}(m_Z^2/m_{H_d}^2(m_Z),1/\tan^2\beta)\right],\\
B_\mu^2\simeq&\L \mu^2+m_{H_u}^2(m_Z)\R m_{H_d}^2(m_Z),\label{BMU}
\end{align}
which are valid for large ${m_{H_d}^2(m_Z)}$ and large $\tan\beta$.
Then we can obtain
\begin{align}\label{}
a_{B}=\f{|r|}{\tan\beta}\ll1,\quad
{\ld_u^2}\simeq \f{I_u}{g(x)\L1-a_B^2\R}\approx \f{I_u}{g(x)},
\end{align}
where we have ignored the RGE corrections to $\mu$ and $B_\mu$, and also neglected
pure gauge terms at the boundary such as $m_{H_d}^2$ and  $m_{H_d}^2$.
Now $\mu\simeq r(x+a_B)I_u\Ld/16\pi^2\sim100$ GeV means $r x\gtrsim {\cal O}(1)$.
Clearly, the desired  parameter values are
  \begin{align}\label{}
|r|\sim1,\,x\sim1 \Rightarrow
\ld_u^2\sim1,\,a_B\simeq \tan^{-1}\beta\sim 0.1.
\end{align}
Next let us turn attention to the fine tuning during the tree loop
cancelation in the complete model. The final $B_\mu$ is
\begin{align}
B_\mu=\Delta_{tree}+\Delta_{loop}=
\frac{-\ld_u\ld_d}{16\pi^2}[\frac{N_F^2\kappa^2F^3}{16\pi^2M^2M_S^2}(\log\frac{\Lambda_{UV}}{M})^2+f_2(\frac{M_S}{M})\frac{F^2}{M^2}]
\end{align}

The tuning is  $\frac{B_\mu}{\Delta_{loop}}$ approximately, which is
about $10\%$ in our scenario. So it is acceptable. But the tuning
increases as $\tan\beta$ gets large, and thus a moderate large
$\tan\beta$ is preferred. Note that although such a tuning is needed
in order to reduce $B_\mu$ considerably, we do not require a
hierarchy between $\ld_u$ and $\ld_d$.

Finally, we summarize the characteristics of our minimal  model as
well as the most natural parameter space
\begin{itemize}
\item We have a small $\mu\simeq 100$ GeV but a large $m_{H_d}$ at  TeV scale,
namely favors the lopsided GSMB scenario.
\item In turn, the mass scale of the singlet $S$ is around the messenger scale
{\it i.e.}, $x\simeq 1$. A high messenger scale is favored in light of
perturbativity of $\ld_u$ and $\ld_d$. But the messenger scale
should low enough if the minimal  model with loop induced tadpole is
viable.

\item The model works at the price of introducing a new source of tuning
around $\sim 10\%$, but not worse than the original
one. Such kind of fine-tuning is acceptable.

\end{itemize}

\subsection{The Supersymmetric CP Problem}

Finally we discuss the supersymmetric CP-problem in our minimal
proposal. The GMSB elegantly solves the flavor problem by virtue of
the flavor blindness of gauge interactions. However, in the MSSM the
CP violation still arises via the flavor-conserving interactions,
$e.g.$, the generation of the electric dipole moment (EDM) of the
fermions via exchanging sparticles. In a class of GMSB models with
vanishing boundary $B_\mu$ and trilinear $A-$terms, $\mu$ can be
rotated to be real via $U(1)_{PQ}-$rotation while the induced phases
of $B_\mu$ and $A$ via gaugino RGE effects are the same as the phase
of the gauginos and can be rotated away through $U(1)_R$ rotation.
So this kind of model is CP
safe~\cite{Rattazzi:1996fb,Gabrielli:1997jp}.

However, in general dynamical models explaining $\mu/B_\mu$ origin,
new CP phases are expected and thus the  phases of $B_\mu$ and $A$
are usually irrelevant to the gaugino phases. So these models could have
the supersymmetric CP problem.
In our model described in Eq.~(\ref{SMODEL}), one cannot take
$M_S$ and $a$ simultaneously real by rephasing $S$, and consequently,
the $\mu/B_\mu$ may have different phases.
To overcome this problem, one may need to consider some more complicated
hidden sector (e.g., see Ref.~\cite{Evans:2010ru}) that is beyond our scope.

\section{Conclusion}

We explored the minimal solution to the $\mu/B_\mu$ problem in gauge
mediation. We introduced a SM singlet field $S$ at about the
messenger scale which couples to the Higgs and messenger fields.
This singlet is nearly supersymmetric and obtains a relatively small
VEV from its radiatively generated tadpole term. Consequently, both
$\mu$ and $B_\mu$ parameters receive the tree-level and one-loop
contributions, which are comparable if the $S$ VEV is small. These
two kinds of contributions allow for a proper cancellation for
$B_{\mu}$ and thus provide a viable Higgs sector for the EWSB.

\section*{Acknowledgments}
This work was supported by the National Natural Science Foundation
of China under grant Nos. 10821504, 10725526 and 10635030, and by
the DOE grant DE-FG03-95-Er-40917.


\end{document}